\documentclass[preprint,10pt,authoryear]{elsarticle}

\usepackage{amsmath}
\usepackage{amssymb}
\usepackage{graphicx}
\usepackage{url}
\usepackage{pdfpages}
\usepackage{todonotes}
\graphicspath{{./Figures/}}

\usepackage{lineno}

\allowdisplaybreaks

\journal{Journal of Theoretical Biology}

\begin{document}

\begin{frontmatter}
\date{}

\title{Infection-acquired versus vaccine-acquired immunity in an SIRWS model}

\author[UoM]{Tiffany~Leung}
\author[UoM]{Barry~D.~Hughes}
\author[SUT]{Federico~Frascoli}
\author[PDI,MCRI]{Patricia~T.~Campbell}
\author[UoM,MSPGH,MCRI]{James~M.~McCaw\corref{cor1}}
\ead{jamesm@unimelb.edu.au}

\cortext[cor1]{Corresponding author}

\address[UoM]{School of Mathematics and Statistics, University of Melbourne, Parkville, Victoria 3010, Australia}
\address[SUT]{Department of Mathematics, Faculty of Science, Engineering and Technology, Swinburne University of Technology, Hawthorn, Victoria 3122, Australia}
\address[PDI]{Peter Doherty Institute for Infection and Immunity, University of Melbourne, Parkville, Victoria 3010, Australia}
\address[MSPGH]{Melbourne School of Population and Global Health, University of Melbourne, Parkville, Victoria 3010, Australia}
\address[MCRI]{Infection and Immunity Research Theme, Murdoch Childrens Research Institute, Royal Children's Hospital, Parkville, Victoria 3052, Australia}
	
\begin{abstract}

Despite high vaccine coverage, pertussis has re-emerged as a public health concern in many countries. One hypothesis posed for re-emergence is the waning of immunity. In some disease systems, the process of waning immunity can be non-linear, involving a complex relationship between the duration of immunity and subsequent boosting of immunity through asymptomatic re-exposure.

We present and analyse a model of infectious disease transmission to examine the interplay between infection and immunity. By allowing the duration of infection-acquired immunity to differ from that of vaccine-acquired immunity, we explore the impact of the difference in durations on long-term disease patterns and prevalence of infection.

Our model demonstrates that vaccination may induce cyclic behaviour, and its ability to reduce the infection prevalence increases with both the duration of infection-acquired immunity and duration of vaccine-acquired immunity. We find that increasing vaccine coverage, while capable of leading to an increase in overall transmission, always results in a reduction in prevalence of primary infections, with epidemic cycles characterised by a longer interepidemic period and taller peaks.

Our results show that the epidemiological patterns of an infectious disease may change considerably when the duration of vaccine-acquired immunity differs from that of infection-acquired immunity. Our study highlights that for any particular disease and associated vaccine, a detailed understanding of the duration of protection and how that duration is influenced by infection prevalence is important as we seek to optimise vaccination strategies.
\end{abstract}

	
\end{frontmatter}

\section{Introduction} \label{Introduction}

The complex dynamics of waning immunity, whether the immunity be acquired through natural infection or through vaccination, play a key role in shaping the epidemiological patterns of an infectious disease. Where immunity is temporary, it wanes over time but may be subsequently boosted upon asymptomatic reexposure, as observed for measles \citep{Whittle1999} and pertussis \citep{Cattaneo1996}. Mathematical models of vaccine-preventable infectious diseases \citep{Glass2003, Aguas2006, Wearing2009, Lavine2011, Campbell2016} are based on the susceptible-infectious-recovered-susceptible (SIRS) framework \citep{Keeling2008}, where every individual in a population is categorised by one of three states based on his or her immune status: susceptible (S) to infection, infected and infectious (I) if they can transmit the infection, and recovered (R) from infection and immune. As immunity wanes, recovered individuals become susceptible to infection again. 

Since the advent of vaccination, mathematical models have been invaluable to vaccination programme design. Key contributions include the introduction of the concept of the basic reproductive ratio $R_0$ \citep{Diekmann1990, Heesterbeek2002, Heffernan2005} and herd immunity thresholds \citep{Fine1993}. Models provide guidance on optimal vaccination strategies \citep{Hethcote2004, Campbell2015, Campbell2017} and reveal insights into the impact of vaccination programmes that may seem counter-intuitive due to nonlinearities in the transmission process.

Vaccines can induce protection against infection, against severe disease, against infectiousness, or a combination of these \citep{Siegrist2008, Preziosi2003a, Preziosi2003}. There are multiple ways through which a vaccine may fail to provide sterilising immunity, such as degree of protection and duration of protection \citep{McLean1993}. They can provide incomplete protection by, for example, reducing the susceptibility by some degree. Furthermore, the protection provided may wane over time.

The reemergence of pertussis over the last two decades has motivated studies on the persistence of immunity to this disease. These studies have found that infections following pertussis vaccination may occur and that these secondary infections may be less severe or asymptomatic, which suggest that protection against infection is shorter than protection against disease \citep{Long1990, Cherry1996, Srugo2000, Wendelboe2005}. The duration of infection-acquired immunity has been estimated to be approximately 60 years \citep{Wearing2009}, compared to a considerably shorter duration of vaccine-acquired immunity of 4--12 years \citep{Wendelboe2005}. It remains an open challenge to determine the vaccine failure mechanism through which a vaccinated individual becomes infected. 

In our study, we focus on the effects of a vaccine with protection that wanes over time. We investigate the influence on infection prevalence as the difference between the durations of vaccine- and infection-acquired immunity is varied. In Section~\ref{sec:model}, we develop our extension to the SIRS-type model by introducing differences between immunity acquired after natural infection and after vaccination. We analyse the model in Section~\ref{sec:methods}. The results are presented in Section~\ref{sec:results}, where we illustrate how infection prevalence changes with vaccine coverage and duration of immunity. Also, we compare the differences in infection prevalence under two different mechanisms through which a vaccine may act to provide protection. A summary of our findings and their relevance are discussed in Section~\ref{sec:discussion}.

\section{The SIRWS model with differences in duration of immunity after natural infection or vaccination} \label{sec:model}

Our model is an extension to the susceptible-infectious-recovered-waning-susceptible (SIRWS) model studied by \cite{Lavine2011} through the inclusion of differences between the duration of infection-acquired immunity and vaccine-acquired immunity. In our extended model, we distinguish between primary and secondary infections to investigate how disease severity, assumed to be lower for secondary infections, may potentially impact case notifications.

Individuals are categorised into one of eight states depending on their immune status. The susceptible population is divided into those who are able to acquire a primary infection ($S_1$) and those who can acquire a secondary infection ($S_2$). Similarly, the infectious population is divided into those with a primary ($I_1$) or secondary ($I_2$) infection. The recovered ($R$) state represents those who have recovered from and are fully immune to infection. They transition to the waning ($W$) state when their immunity has waned sufficiently. From there, they can become susceptible to secondary infections (transition to $S_2$), or have their immunity boosted upon reexposure (return to $R$).

Vaccination is represented in the model by a fraction of newborns entering the population through the vaccinated ($V$) state, where they are temporarily fully protected against infection. As vaccine-acquired immunity wanes, vaccinated individuals enter the waning vaccine-acquired immunity ($W_v$) state. Over time, they either lose all immunity and become susceptible to primary infection ($S_1$), or their immunity may be boosted. Under our first model for how vaccination provides protection, upon a boosting of vaccine-acquired immunity, the vaccinated individual enters the $R$ compartment, where subsequent waning of immunity would lead them to $S_2$. Should they then become infectious, they would be infected with a secondary infection (with lower disease severity)---bypassing a primary infection. The model is described by the following system of differential equations for the proportions of the population in each state:
\begin{subequations}
	\label{sirws}
	\begin{align}
	\frac{dS_1}{dt} &= (1-p) \mu - \lambda S_1 + 2 \kappa_v W_v - \mu S_1 \, , \\
	\frac{dI_1}{dt} &= \lambda S_1 - \gamma I_1 - \mu I_1 \, , \\
	\frac{dR}{dt} &= \gamma I_1 + \gamma I_2 - 2 \kappa_n R + \nu \lambda W + \nu \lambda W_v - \mu R \, , \label{subeq:drdt} \\
	\frac{dW}{dt} &= 2 \kappa_n R - 2 \kappa_n W - \nu \lambda W - \mu W \, , \\
	\frac{dS_2}{dt} &= 2 \kappa_n W - \lambda S_2 - \mu S_2 \, , \\
	\frac{dI_2}{dt} &= \lambda S_2 - \gamma I_2 - \mu I_2 \, , \\
	\frac{dV}{dt} &= p \mu - 2 \kappa_v V - \mu V \, , \label{subeq:dvdt} \\
	\frac{dW_v}{dt} &= 2 \kappa_v V - 2 \kappa_v W_v - \nu \lambda W_v - \mu W_v \, .
	\end{align}
\end{subequations}

The rate at which the proportion of susceptible individuals acquires an infection and becomes infectious is the force of infection, denoted by $\lambda = \beta (I_1 + I_2)$, where $\beta$ is the transmission coefficient. Secondary infections are considered equally infectious as primary infections, both with an average recovery time of $1/\gamma$. Births and deaths occur at an equal rate $\mu$, so that the population size is constant, and disease-induced mortality is ignored. The proportion of births entering the vaccinated state is denoted by $p$.

The waning of natural immunity is modelled as a two-step process from $R$ to $S_2$ via the state $W$. We use the method of stages \citep{Lloyd2001} to model this transition with the parameter $\kappa_n$. In the absence of boosting, the transition time from $R$ to $S_2$ is $2/(2\kappa_n + \mu)$, so that for $\mu \ll \kappa_n$, the average duration of immunity is approximately $1/\kappa_n$, the same mean duration as for the standard SIRS model with immunity waning rate $\kappa_n$. The waning of vaccine-acquired immunity (from $V$ to $S_1$ at rate $\kappa_v$) is similarly defined. Immune boosting occurs at a rate $\nu \lambda$, where $\nu$ is the strength of immune boosting relative to the force of infection ($\nu > 0$). Allowing $\nu > 1$ implies that boosting requires lower exposure than for a transmissible infection. See Table~\ref{table:parameters} for a description of the state variables and parameters.

\begin{figure}[tb] \centering
	\includegraphics[width=0.75\textwidth]{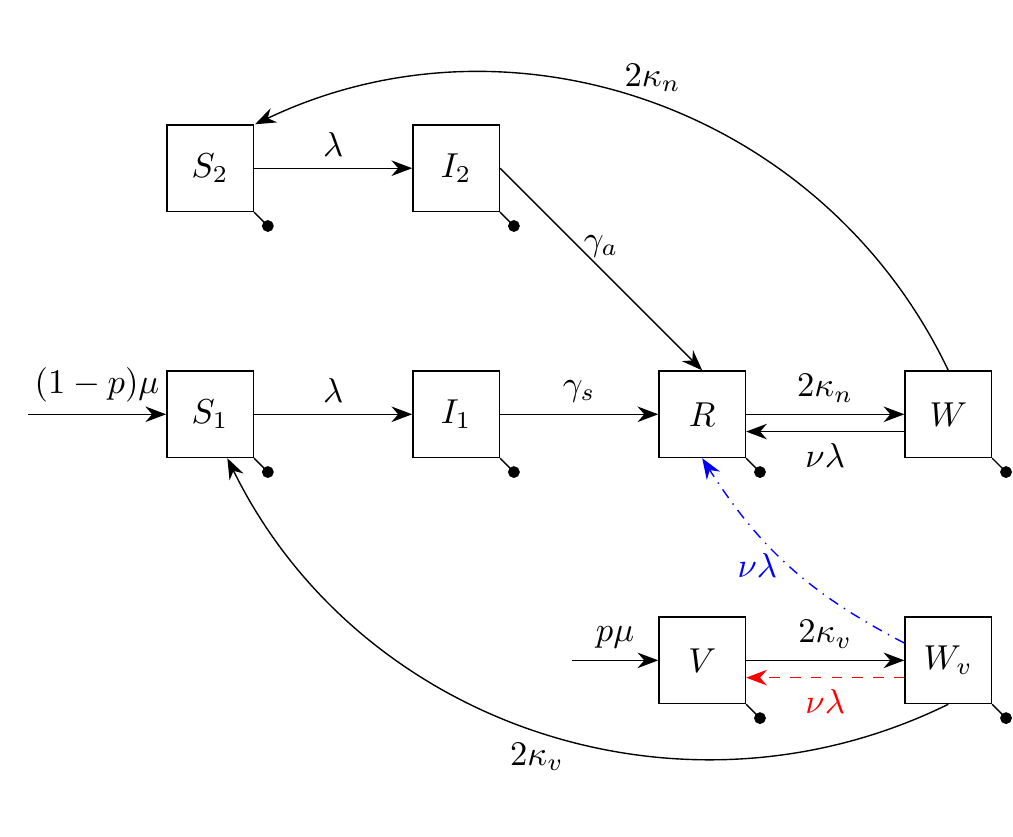}
	\caption{(Colour online) A diagram of the extended SIRWS model with infection-acquired and vaccine-acquired immunity. The population is divided into eight different states, represented by boxes. A boosting of vaccine-acquired immunity may either bypass the primary infection (blue dash-dotted line) or delay the primary infection (red dashed line). The death rate $\mu$ is denoted by a bullet.}
	\label{fig:model}
\end{figure}

Under our second model for vaccination protection, we consider the case where the boosting of vaccine-acquired immunity delays (rather than bypasses) a primary infection. This can be examined by reformulating Equations~\eqref{subeq:drdt} and \eqref{subeq:dvdt} into 
\begin{align*}
\frac{dR}{dt} &= \gamma I_1 + \gamma I_2 - 2 \kappa_n R + \nu \lambda W - \mu R \, , \label{subeq:drdt-delay} \\
\frac{dV}{dt} &= p \mu + \nu \lambda W_v - 2 \kappa_v V - \mu V \, .
\end{align*}
The flow diagrams for each of these vaccine-acquired immune boosting mechanisms are illustrated in Figure~\ref{fig:model}.

\begin{table}[tb] \centering \footnotesize{
		\begin{tabular}{p{1.2cm} p{8.6cm} p{2cm}}
			\hline
			Parameter & Description & Default \\
			\hline
			$p$ & vaccinated proportion of the population & (varies) \\
			$\beta$ & transmission coefficient & 260 y$^{-1}$p$^{-1}$  \\
			$\gamma$ & recovery rate  & 17 y$^{-1}$ \\
			$\nu$ & strength of immune boosting relative to the force of infection & 3 \\
			$\kappa_n$ & rate of loss of infection-acquired immunity & (varies) y$^{-1}$ \\
			$\kappa_v$ & rate of loss of vaccine-acquired immunity & (varies) y$^{-1}$ \\
			$\mu$ & birth and death rate & 1/80 y$^{-1}$ \\ \\
			\hline
			State & Description \\
			\hline
			$S_1$ & susceptible to primary infection \\
			$I_1$ & infectious with primary infection \\
			$R$ & recovered from an infection and fully immune \\
			$W$ & recovered from an infection with waning immunity \\
			$S_2$ & susceptible to a secondary infection \\
			$I_2$ & infectious with secondary infection \\
			$V$ & vaccinated and fully immune \\
			$W_v$ & vaccinated with waning vaccine-acquired immunity \\
			\hline
	\end{tabular} }
	\caption{Descriptions of the parameters and state variables. (y = years; p = person)} \label{table:parameters}
\end{table}

Our extended SIRWS model in Equation~\eqref{sirws} collapses to the SIRWS model studied by \cite{Lavine2011} under the following substitutions:
\begin{align*}
\mathbb{S} &= S_1 + S_2 \, , \\
\mathbb{I} &= I_1 + I_2 \, , \\
\mathbb{R} &= R + V \, , \\
\mathbb{W} &= W + W_v \, , \\
\kappa &= \kappa_n = \kappa_v \, .
\end{align*}
\cite{Dafilis2012} have shown that the inclusion of immune boosting allows the SIRWS model to produce limit cycles in the absence of vaccination---dynamics that are qualitatively different to the standard SIRS model. Further, \cite{Leung2016} observed that limit cycles were absent in the SIRWS model when immune boosting occurred at a rate less than the force of infection (that is, $\nu < 1$). Additionally, for limiting values of waning immunity ($\kappa \rightarrow 0$ and $\kappa \rightarrow \infty$), the extended model reduces to the standard SIR and SIS model respectively.

\section{Methods} \label{sec:methods}

The extended SIRWS model described by Equation~\eqref{sirws} was numerically integrated in MATLAB 2014b's \texttt{ode45} ODE solver (The MathWorks Inc., Natick, Massachusetts) with absolute and relative tolerance both set to $1 \times 10^{-8}$. The system was run for 1300 years with the first 700 years discarded as transient. Initial conditions were set to be close to the endemic equilibrium of the SIRWS model \citep{Lavine2011} in the absence of vaccination: 
\[ (S_1, I_1, R, W) = (0.065, 0.001, 0.760, 0.174) \] 
and all other states were set to 0.

The transmission coefficient and recovery rate ($\beta$ and $\gamma$ respectively) were taken from \cite{Lavine2011} in their study of pertussis. Other parameter values were chosen so that the model produces periodic cycles in the absence of vaccination. As per \cite{Leung2016}, the relative strength of immune boosting $\nu$ was set to 3. The average life expectancy ($1/\mu$) was set to 80 years. The default model parameters are detailed in Table~\ref{table:parameters}. 

As we examine the influence of duration of immunity on the epidemiological patterns of disease, we introduce $T_n = 1/\kappa_n$ and $T_v = 1/\kappa_v$ to denote the average duration of infection- and vaccine-acquired immunity respectively (in the absence of immune boosting). We allow $T_n$ to range over 5--80 years, which covers suggested estimates for the duration of immunity following a pertussis infection. In \cite{Lavine2011}'s study of vaccination using the SIRWS model it was assumed that $T_v$ and $T_n$ were equal. In our study, we consider the case where the duration of vaccine-acquired immunity is less than or equal to that of infection-acquired immunity. Accordingly, the duration of vaccine-acquired immunity $T_v$ ranges from 2 years to as high as the corresponding $T_n$.

To generate diagrams of the mean infection prevalence as a function of the vaccinated proportion ($p$), duration of vaccine-acquired immunity ($T_v$), and duration of infection-acquired immunity ($T_n$), the values of $T_v$ were chosen as follows. For $T_n \leq 30$ years, $T_v$ was chosen by dividing $T_n$ into five equal intervals. For $T_n > 30$ years, $T_v$ was set to intervals of 10 years up to the respective $T_n$.

The average infection prevalence was calculated as the mean of the time series of the infectious proportion ($I_1 + I_2$) over 40 peaks in the presence of periodic cycles. Where periodic cycles were absent, the average infection prevalence was set to be equal to the numerically obtained endemic equilibrium of the system. The distribution of the population over the eight different states was similarly calculated. Bifurcation analyses were performed in XPPAUT 8.0 \citep{Ermentrout2002} with an adaptive step size Runge--Kutta integrator.

\section{Results} \label{sec:results}

In our extended SIRWS model, we explore two mechanisms of vaccine protection, where vaccine-acquired immune boosting may either bypass or delay the primary infection. We find that the qualitative behaviour of the model under both mechanisms is similar. Hence, we present the results for the mechanism where vaccine-acquired immune boosting bypasses the primary infection. We provide the corresponding figures for the alternative mechanism in the Appendix.

\subsection{Influence of vaccination to lower infection prevalence is more sensitive to the duration of infection-acquired immunity than of vaccine-acquired immunity.} \label{sec:meanofi}

\begin{figure} \centering 
	\includegraphics[width=\textwidth]{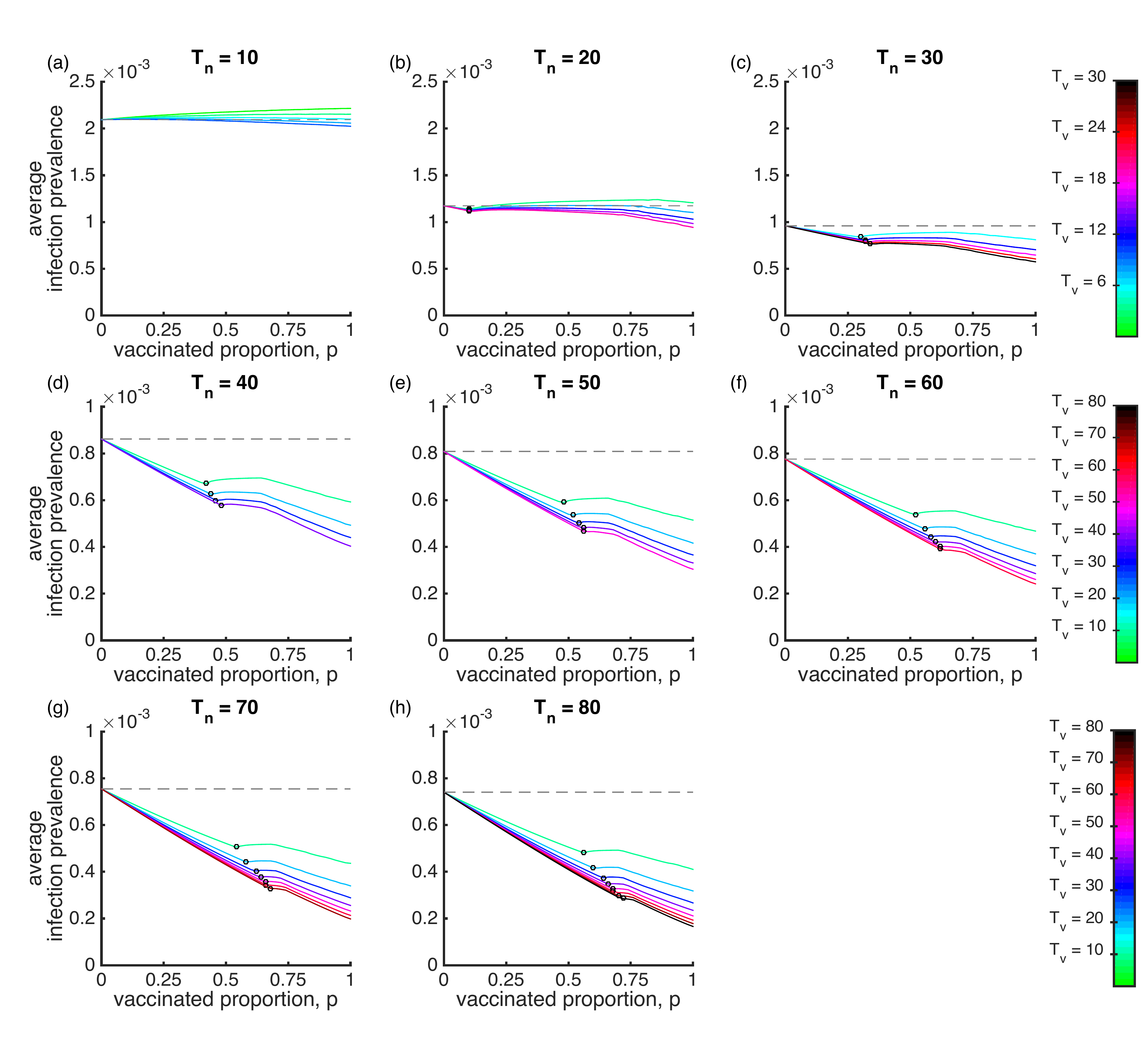}
	\caption{(Colour online) The mean infection prevalence for varying vaccinated proportion, duration of infection-acquired immunity (plots) and duration of vaccine-acquired immunity (colours). The gray dashed line indicates the mean infection prevalence in the absence of vaccination ($p = 0$). The black circle on each line denotes the Hopf bifurcation point that generates periodic cycles.}
	\label{fig:avg_i_prevalence}
\end{figure}

We first examine the impact of vaccine coverage and waning immunity on the mean infection prevalence ($I_1 + I_2$). Figure~\ref{fig:avg_i_prevalence} shows the mean infection prevalence for eight different durations of infection-acquired immunity, $T_n$, and varying durations of vaccine-acquired immunity, $T_v$. For relatively short infection-acquired immunity $T_n = (10, 20, 30)$, waning immunity is a transient process relative to the demographic time scale. For long-lived infection-acquired immunity ($T_n \geq 40$), the duration of immunity becomes comparable to demographic time scales. The gray dashed line marks the mean infection prevalence in the absence of vaccination. As vaccination is introduced ($p > 0$), the infection prevalence is characterised by a stable endemic equilibrium (Figure~\ref{fig:avg_i_prevalence}b--h). As $p$ increases, the endemic equilibrium bifurcates via a Hopf bifurcation---denoted by the black circle---into stable periodic cycles, and the mean infection prevalence is calculated from a time series over 40 cycles. For $T_n = 10$ (Figure~\ref{fig:avg_i_prevalence}a), the Hopf bifurcation occurs at a non-biological region ($p < 0$) and thus is not shown.

The ability of a vaccine to reduce infection prevalence increases with the duration of both infection- and vaccine-acquired immunity. When natural immunity is relatively short ($T_n \leq 30$), the reduction in infection prevalence is more sensitive to the duration of infection-acquired immunity (Figure~\ref{fig:avg_i_prevalence}a--c). For long-lived natural immunity ($T_n \geq 40$), the duration of vaccine-acquired immunity has a larger influence on reducing the infection prevalence (Figure~\ref{fig:avg_i_prevalence}d--h). The mean infection prevalence generally declines as vaccine coverage increases. However, when natural immunity is relatively short-lived, the mean infection prevalence may reach a level greater than that without vaccination, as observed in Figure~\ref{fig:avg_i_prevalence}a--b. We examine this surprising result in detail in Section~\ref{sec:popn_distribution}. 

Similar behaviour in infection prevalence is observed when the boosting of vaccine-acquired immunity delays (rather than bypasses) a primary infection (see \ref{app:peaks_ac}), or when the demographic timescale changes (see \ref{app:life_expectancy}). In the former case, while the behaviours of our extended SIRWS model under two different mechanisms of vaccine protection are generally similar, the vaccinated proportion has a bigger influence on decreasing the mean infection prevalence in the system where the boosting of vaccine-acquired immunity bypasses the primary infection. In the latter case, extending the average life expectancy would allow the process of waning immunity to be more transient (or alternatively, more comparable if average life expectancy decreases). For a given duration of natural immunity and a vaccine with a fixed duration of protection, the ability of a vaccine to further reduce infection decreases as the average life expectancy increases.

In the context of pertussis, the estimated duration of immunity after a pertussis infection ranges from 3.5 years \citep{Versteegh2002} to lifelong \citep{Gordon1951} (though improved diagnostic testing over time may be a contributing factor to the striking difference). In turn, the ability of a vaccine to lower infection prevalence also varies considerably. For example, for a vaccine lasting an average of say, 10 years, the model shows a greater reduction in infection when the duration of natural immunity is 60 years compared to when it is 10 years. 

\subsection{The mean primary infection prevalence decreases as vaccine coverage increases, but the mean total infection prevalence can increase} \label{sec:popn_distribution}

Here we examine the surprising result from Section~\ref{sec:meanofi} that the mean total infection prevalence may reach a level greater than that without vaccination as vaccine coverage increases. We remind the reader that the total infection prevalence is the sum of two infectious states, $I_1 + I_2$. To look further into the behaviour of each of these infectious states, we calculated the mean proportion of each immune status state (the eight compartments of the extended SIRWS model). 

Although the mean total infection prevalence may increase with vaccine coverage, the mean primary infection prevalence always decreases with increasing duration of vaccine-acquired immunity and vaccine coverage. However, this decrease in primary infection may be countered by an increase in transmission from secondary infections.

\begin{figure} \centering
	\includegraphics[width=\textwidth]{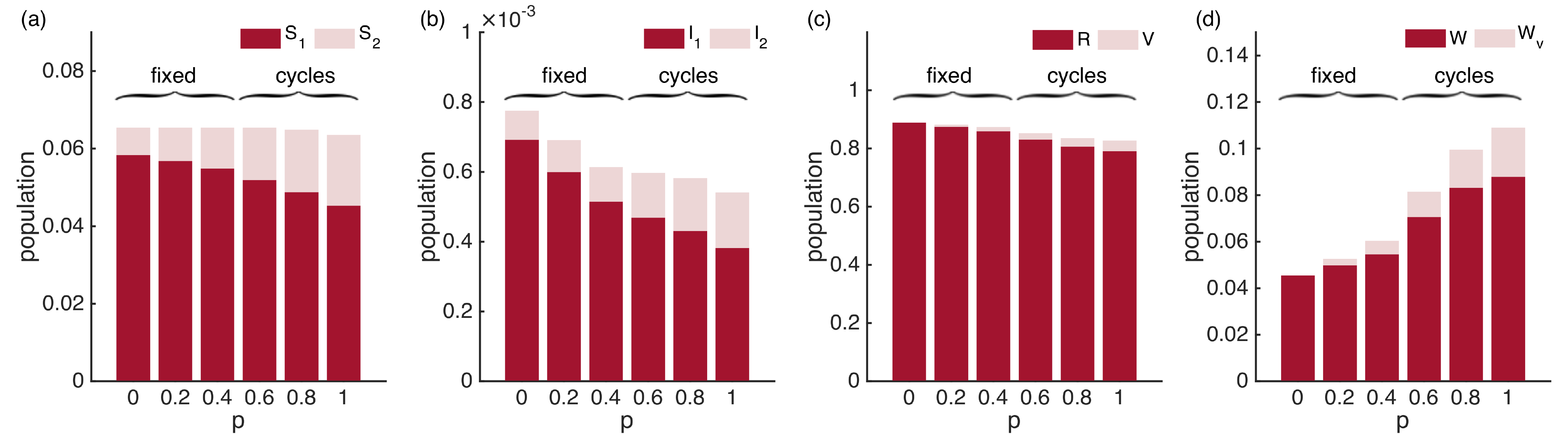}
	\includegraphics[width=\textwidth]{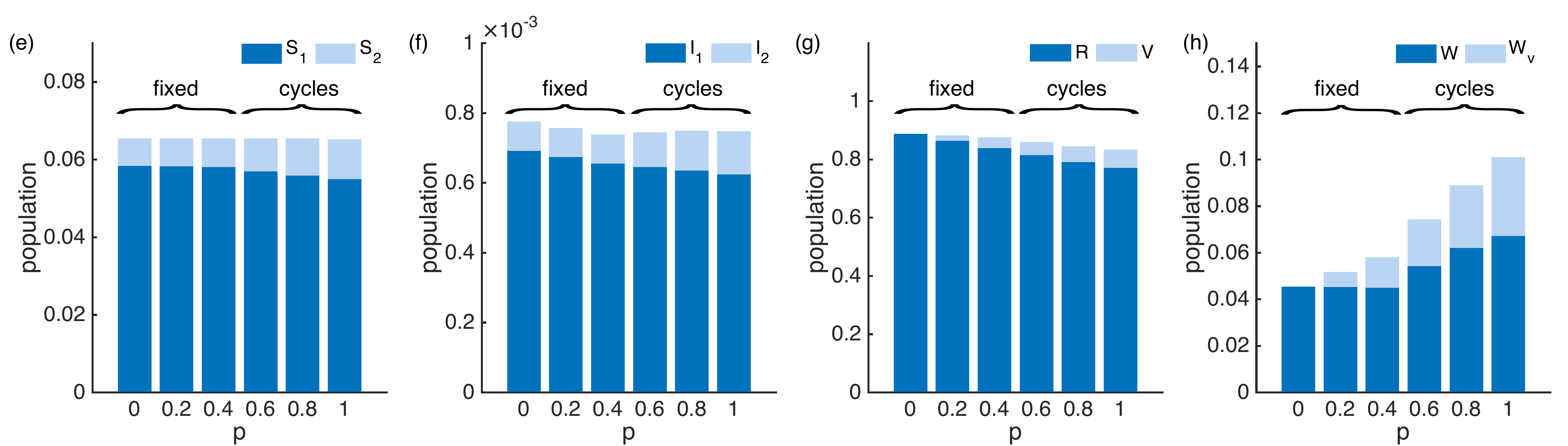}
	\caption{(Colour online) The mean of the eight states for the extended SIRWS model in which a boosting of vaccine-acquired immunity bypasses (top) or delays (bottom) a primary infection. ($T_n = 60$ years; $T_v = 6$ years)}
	\label{fig:bar_popn}
\end{figure}

Figure~\ref{fig:bar_popn} shows the distribution of the population for the extended SIRWS model under both mechanisms of vaccination protection: where the boosting of vaccine-acquired immunity bypasses (top) or delays (bottom) a primary infection. Indeed, Figure~\ref{fig:bar_popn}b and d show that the mean primary infection prevalence is always decreasing as vaccine coverage increases, regardless of whether the mean was calculated over a stable fixed point or periodic cycles. Figure~\ref{fig:bar_popn}f shows that an increasing mean secondary infection prevalence may contribute to an increase in mean total infection prevalence, despite a reducing proportion of primary infections.

In the context of an infectious disease, a primary infection may be associated with severe disease, and a secondary infection be considered mild or asymptomatic. Vaccination could plausibly act by protecting the individual from disease (if not by blocking the infection altogether). This may potentially lead to a reduction in morbidity and consequently fewer case notifications.

\subsection{The peak in infection prevalence decreases with increasing duration of infection-acquired immunity}

\begin{figure}
	\includegraphics[width=\textwidth]{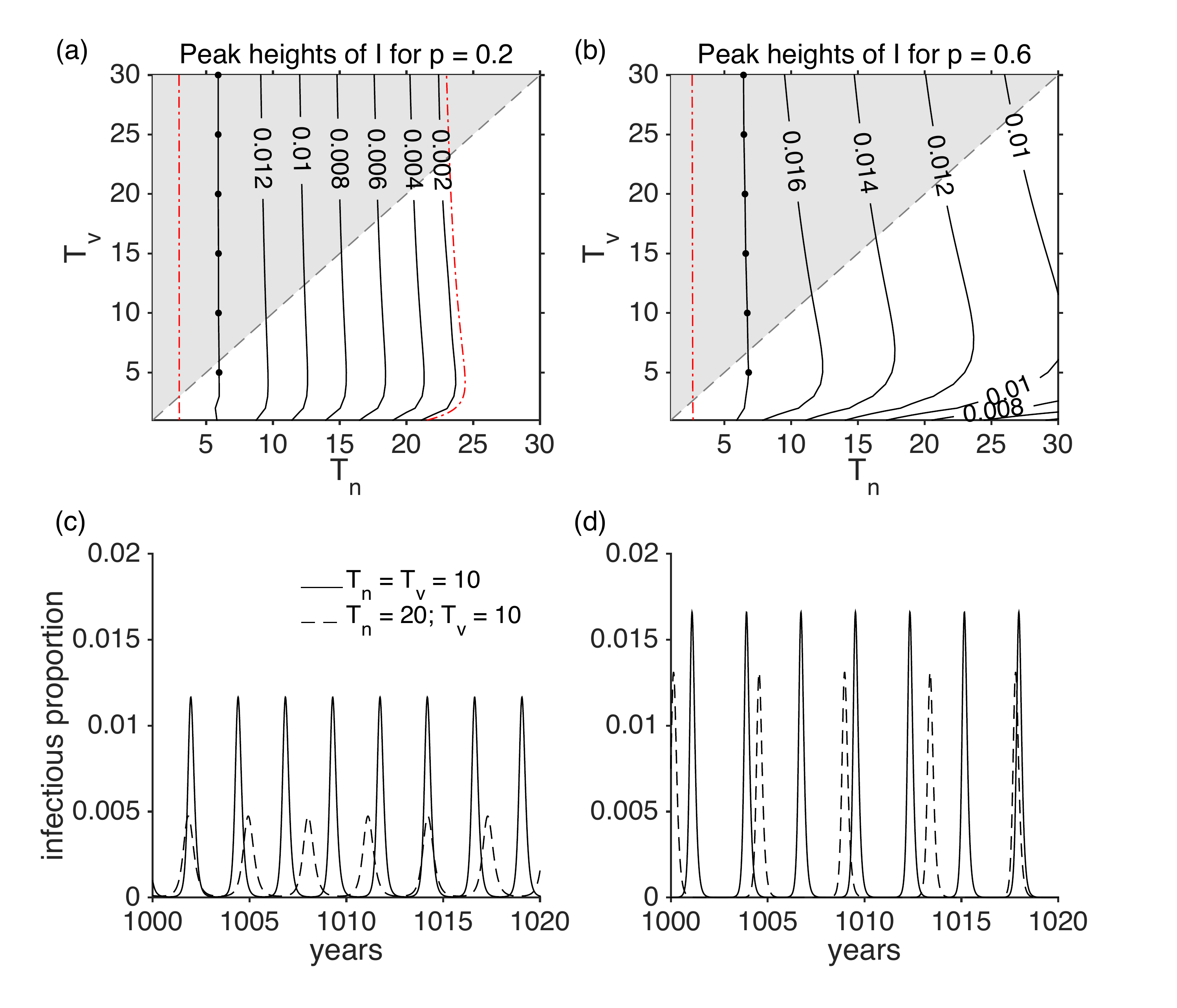}
	\caption{(Colour online) The infection prevalence peaks for varying durations of infection-acquired immunity ($T_n$) and durations of vaccine-acquired immunity ($T_v$), illustrated through contour plots (a--b) and time series (c--d). The shaded area shows where $T_v > T_n$, and the gray dashed line shows where $T_v = T_n$. Between the Hopf bifurcation branches (shown by red dash-dotted lines) are contour lines of the peak heights of the infectious proportion. The line with filled circles indicates the points in parameter space with the maximum peak height of the periodic cycles.}
	\label{fig:peaks}
\end{figure}

In addition to the mean infection prevalence, the peak height and interepidemic period of the periodic cycles are important measures to consider. The maximum proportion of infectious cases in the presence of periodic cycles is shown in Figure~\ref{fig:peaks}a--b for a low and moderate vaccinated proportion at $p = 0.2$ (a) and $p = 0.6$ (b) respectively. The shaded region indicates when vaccine-acquired immunity lasts longer than infection-acquired immunity, and the dashed line shows where their durations are the same. For vaccine-preventable childhood infectious diseases such as measles or pertussis, the biologically plausible parameter space is where $T_n \geq T_v$ (unshaded area). 

The peak heights of the periodic cycles are best described by considering the duration of natural immunity transitioning from long-lived to short-lived. The long-term dynamics of the SIRWS system as $T_n \rightarrow \infty$ are characterised by a fixed point attractor, similar to the SIR model. As $T_n$ decreases (moving from right to left in Figure~\ref{fig:peaks}a), the peak heights start from being equal to the stable fixed point (with an amplitude of 0 in the absence of periodic cycles) until a Hopf bifurcation (shown by a red dash-dotted line) gives rise to stable periodic cycles. For lower values of $T_n$ still, the periodic cycles lose stability through a second Hopf bifurcation, and the dynamics are characterised by a stable fixed point once again. After the first Hopf bifurcation (around $T_n \approx 23$) as $T_n$ decreases, for a fixed $T_v$, the peak height of the periodic cycles monotonically grows until it reaches a maximum (the line of maximum peak heights for a given $T_v$ is shown by the line with filled circles). After reaching a maximum, the peak height decreases again. For clarity, the contour lines for these decreasing peak heights (between the line with filled circles and the Hopf bifurcation line at $T_n \approx 2$) have been omitted. The behaviour of the peak heights for higher vaccine coverage at $p = 0.6$ (Figure~\ref{fig:peaks}b) is similar. However, over a decreasing $T_n$, the Hopf bifurcation giving rise to periodic cycles occurs at a value of $T_n$ that is higher than 30 (and is thus not shown).

The duration of vaccine-acquired immunity has a small effect on the epidemic peak height relative to the duration of infection-acquired immunity, as shown in Figure~\ref{fig:peaks}a--b. There is a noticeable decrease in the maximum proportion of infectious cases as duration of infection-acquired immunity increases. The difference in peak height is illustrated through the time series for two different combinations of immunity durations at $p = 0.2$ (Figure~\ref{fig:peaks}c) and $p = 0.6$ (Figure~\ref{fig:peaks}d). Additionally, a growing interepidemic period with taller peak height is observed by increasing the vaccinated proportion (compare Figure~\ref{fig:peaks}c to Figure~\ref{fig:peaks}d). These epidemiological patterns are also observed when the boosting of vaccine-acquired immunity delays (rather than bypasses) a primary infection (See \ref{app:peaks_ac}). Taken together, increasing vaccine coverage leads to fewer primary infections (Figure~\ref{fig:bar_popn}b), with periodic cycles characterised by taller peaks (Figure~\ref{fig:peaks}a--b) and a longer interepidemic period (Figure~\ref{fig:peaks}c--d). 

\begin{figure} \centering
	\includegraphics[width=\textwidth]{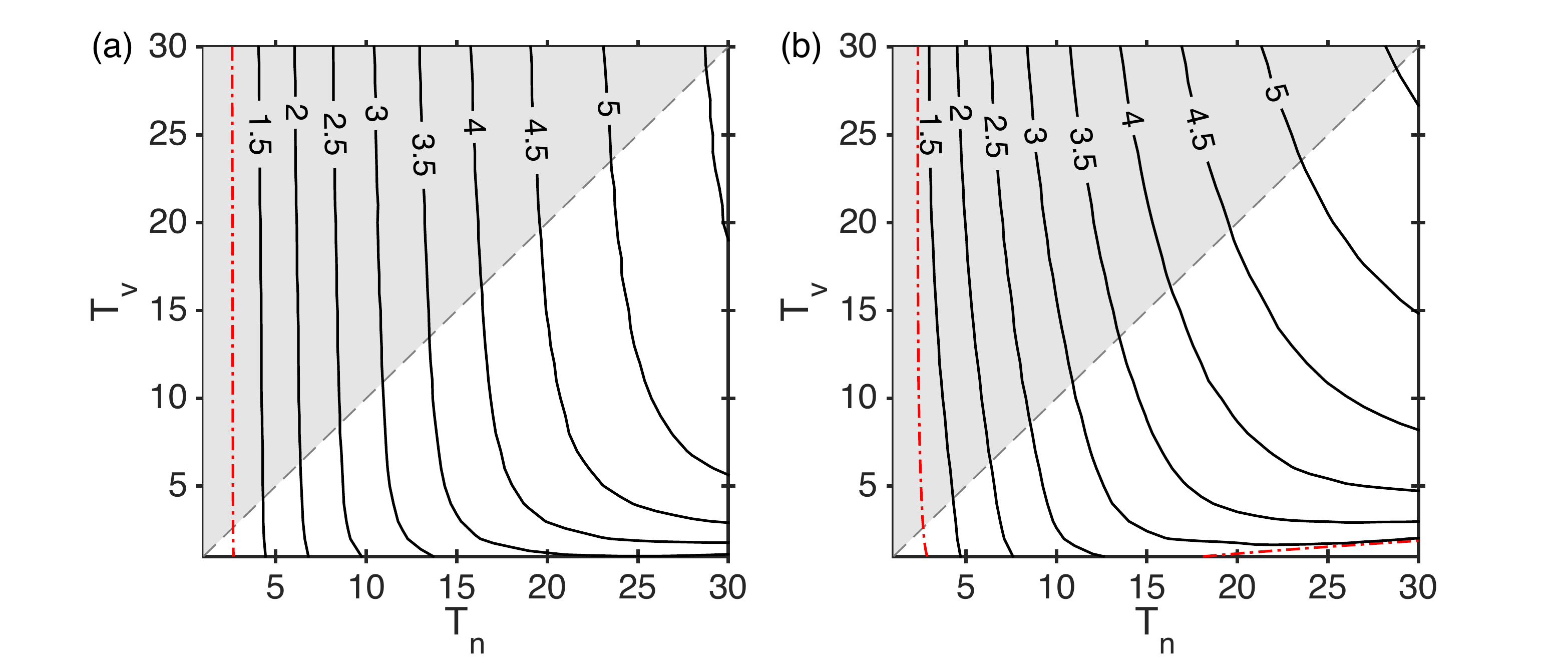}
	\caption{(Colour online) The interepidemic period of the SIRWS model at $p = 0.6$ where the boosting of vaccine-acquired immunity bypasses (a) or delays (b) a primary infection. The gray dashed line indicates where the duration of infection-acquired and vaccine-acquired immunity are the same. The red dash-dotted line represents the branch of Hopf bifurcations, where the period is approximately 1.}
	\label{fig:period}
\end{figure}

The interepidemic period is also more sensitive to the duration of infection-acquired immunity than the duration of vaccine-acquired immunity. It grows with increasing $T_n$ and $T_v$, as illustrated in Figure~\ref{fig:period} at $p = 0.6$, where the boosting of vaccine-acquired immunity either bypasses (a) or delays (b) a primary infection. In the presence of periodic cycles for a small vaccinated proportion, the outbreaks are characterised by lower peaks (Figure~\ref{fig:peaks}a) with shorter interepidemic periods. In contrast, the outbreaks are described by taller peaks and longer interepidemic periods under a higher vaccinated proportion. 

In the public health context, the epidemiological patterns of an infectious disease may change considerably when the duration of vaccine-acquired immunity differs from the duration of infection-acquired immunity. Although the mean infection prevalence generally decreases with increasing vaccine coverage (Figure~\ref{fig:avg_i_prevalence}), reducing the cumulative public health burden may concentrate the burden on public health services into short intervals.

\section{Discussion} \label{sec:discussion}

We have shown with our extended SIRWS model that allowing the duration of infection-acquired immunity to differ from the duration of vaccine-acquired immunity can have substantial impact on infection prevalence. The duration of infection-acquired immunity plays a large role in determining the ability of vaccination to reduce infection prevalence. While the advent of periodic cycles can increase the average infection prevalence as vaccine coverage increases, average primary infection prevalence always falls with increasing vaccination. The infection peaks decrease in height as the duration of infection-acquired immunity increases.

Our results show that the epidemiological patterns of disease can change considerably with variation in the duration of infection-acquired and vaccine-acquired immunity. The presence of vaccination may induce large-amplitude oscillations. We have demonstrated that by accounting for longer infection-acquired immunity than vaccine-acquired immunity, the maximum proportion of the population that is infectious decreases, and the amplitude of the oscillations becomes smaller. The period of the oscillations grows with the duration of immunity (both vaccine-acquired and infection-acquired) to decrease the average infection prevalence. By extending the SIRWS model, we find that increasing vaccine coverage decreases the proportion of primary infections, though this reduction may be countered by an increase in transmission and secondary infections.

Our results complement those of \cite{Heffernan2009}, who also found that the interplay between immunity and vaccination can induce large-amplitude oscillations. Our finding that decreased primary infection may not have a substantial impact on the overall infection prevalence due to secondary infections is consistent with that of \cite{Aguas2006}. Others \citep{Aguas2006, Campbell2016} have included immune boosting into their transmission models. However, different from \cite{Aguas2006} who modelled the boosting of immunity as an infectious process, our interpretation of boosting as a non-infectious process allows immune boosting to be more easily triggered than a transmissible infection ($\nu > 1$). The results from the SIRWS model depend critically on this assumption; without it the model is unable to generate sustained oscillations ($\nu < 1$) \citep{Leung2016}.

In the situation where primary infections are associated with disease and secondary infections are mild or asymptomatic, as observed for pertussis \citep{Mertsola1983, Long1990}, increasing vaccine coverage reduces the burden of disease. However, the peaks associated with the oscillations could stretch public health services over short intervals.

Vaccines may also reduce an individual's ability to acquire or transmit a subsequent infection. We have not explored the potential impact of reduced susceptibility or reduced infectiousness on the epidemiological patterns of disease. Indeed, a study by \cite{Preziosi2003} found that pertussis cases from vaccinated individuals were less contagious. How these reductions may influence the composition of the infectious population and epidemiological patterns remains an open question. Our study highlights the importance of obtaining a better understanding of the persistence of immunity. Untangling the mechanisms responsible for protection against infection, against disease and against transmission remains a challenge.

\section*{Declaration of interest}
Conflicts of interest: none.

\section*{Author contribution}
TL, BH and JM conceived the study. TL, BH, PC and JM developed the mathematical models. TL performed the analysis with assistance from FF, BH and JM. TL drafted the manuscript. All authors read and approved the final manuscript.

\section*{Acknowledgements}
Tiffany Leung is supported by a Melbourne International Research Scholarship from the University of Melbourne and a National Health and Medical Research Council (NHMRC) funded Centre for Research Excellence in Infectious Diseases Modelling to Inform Public Health Policy (1078068).

\newpage
\appendix

\section{Corresponding figures of the SIRWS model where the boosting of vaccine-acquired immunity delays a primary infection} \label{app:peaks_ac}

\begin{figure}[hb] \centering
	\includegraphics[width=\textwidth]{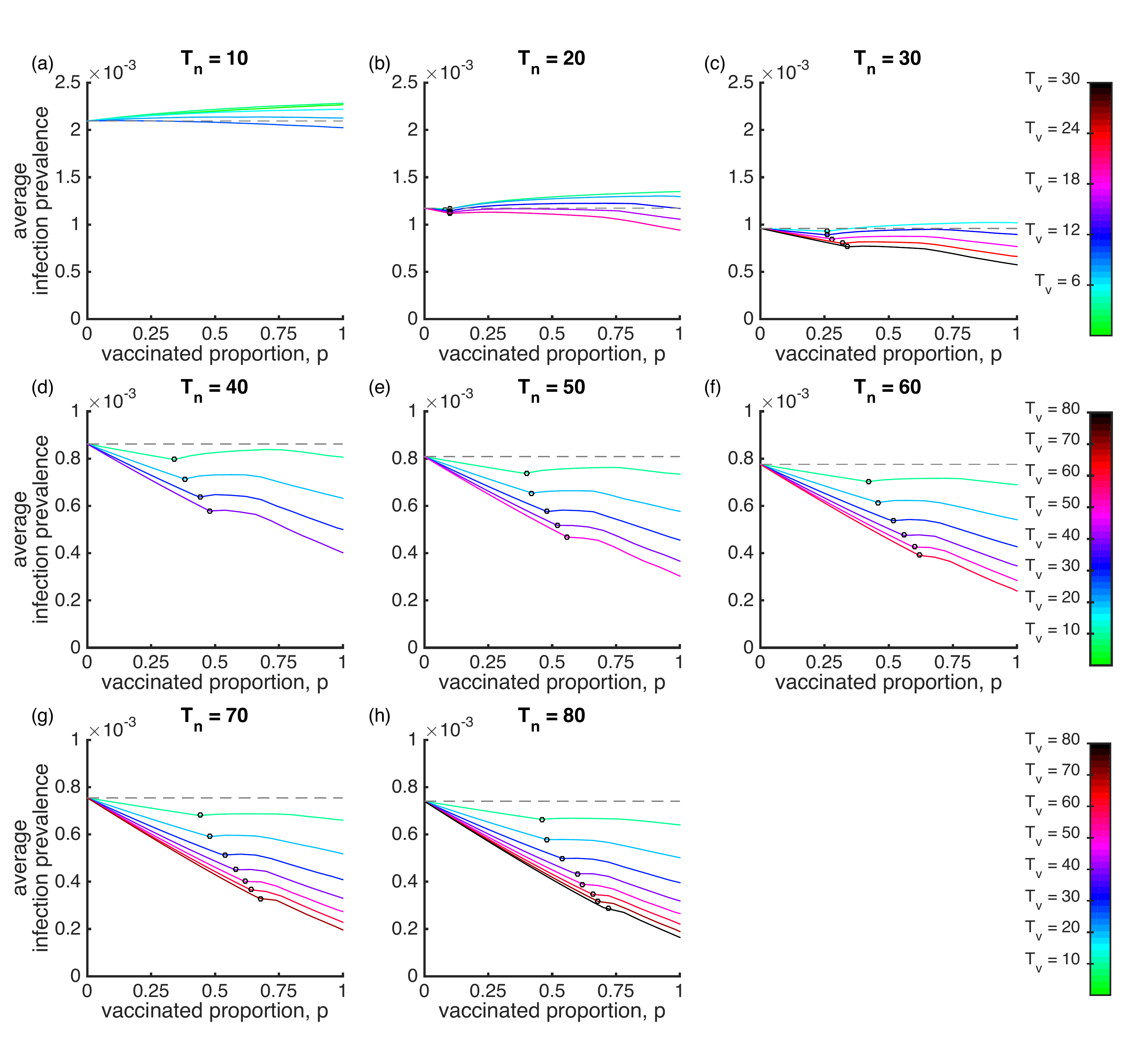}
	\caption{(Colour online) The average infection prevalence for varying vaccinated proportion, duration of infection-acquired immunity (plots) and duration of vaccine-acquired immunity (colours), where the boosting of vaccine-acquired immunity delays primary infection. The gray dashed line indicates the mean infection prevalence in the absence of vaccination ($p = 0$). The black circles denote the Hopf bifurcation points that generate periodic cycles.}
	\label{fig:avg_i_prevalence_ac}
\end{figure}

\begin{figure}
	\includegraphics[width=\textwidth]{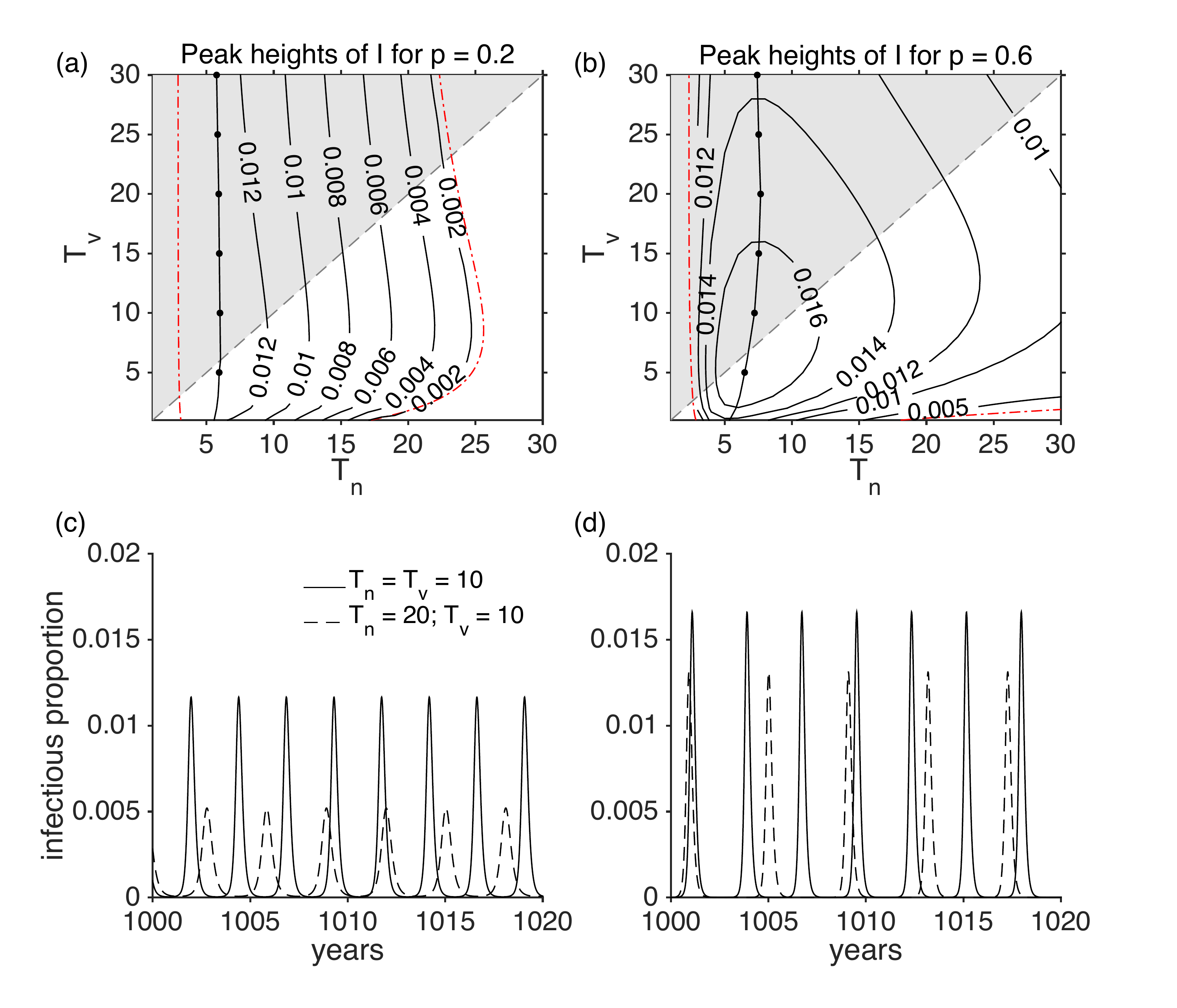} 
	\caption{(Colour online) The infection prevalence peaks for varying durations of infection-acquired immunity ($T_n$) and durations of vaccine-acquired immunity ($T_v$), where the boosting of vaccine-acquired immunity \textit{delays} a primary infection. Contour plots of the peak heights of the infectious proportion are displayed for $p = 0.2$ (a) and $p = 0.6$ (b). The time series are shown for $p = 0.2$ (c) and $p = 0.6$ (d). The shaded area indicates where $T_v > T_n$, and the gray dashed line marks where $T_v = T_n$. The Hopf bifurcation lines are shown by red dash-dotted lines. The line with filled circles indicates the points in parameter space with the maximum peak height of the periodic cycles.} \label{fig:peaks_ac}
\end{figure}

In this appendix, we repeat the numerical calculations for the extended SIRWS model where the boosting of vaccine-acquired immunity delays a primary infection. Figure~\ref{fig:avg_i_prevalence_ac} shows the mean infection prevalence as a function of the vaccinated proportion for different values of the duration of infection-acquired immunity. Figure~\ref{fig:peaks_ac} displays the peaks of the infection prevalence for varying durations of immunity.

\section{Mean infection prevalence for different average life expectancy} \label{app:life_expectancy}

\begin{figure}[hb] \centering
	\includegraphics[width=\textwidth]{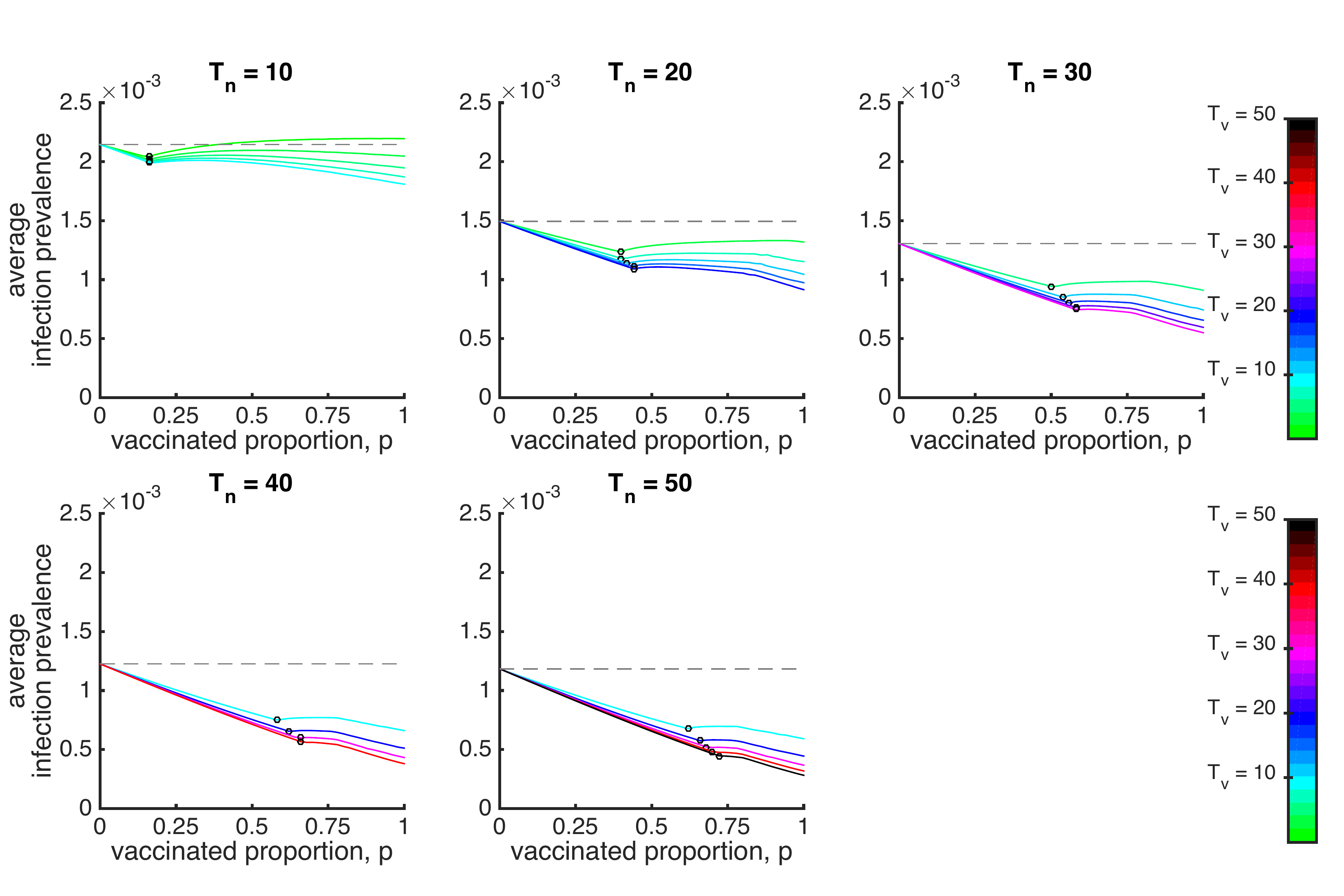}
	\caption{(Colour online) The mean infection prevalence for varying vaccinated proportion, duration of infection-acquired immunity (plots) and duration of vaccine-acquired immunity (colours). The gray dashed line indicates the mean infection prevalence in the absence of vaccination ($p = 0$). The black circle on each line denotes the Hopf bifurcation point that generates periodic cycles. ($1/\mu = 50$ years)} \label{fig:avg_i_prevalence_mu50}
\end{figure}

\begin{figure} \centering
	\includegraphics[width=\textwidth]{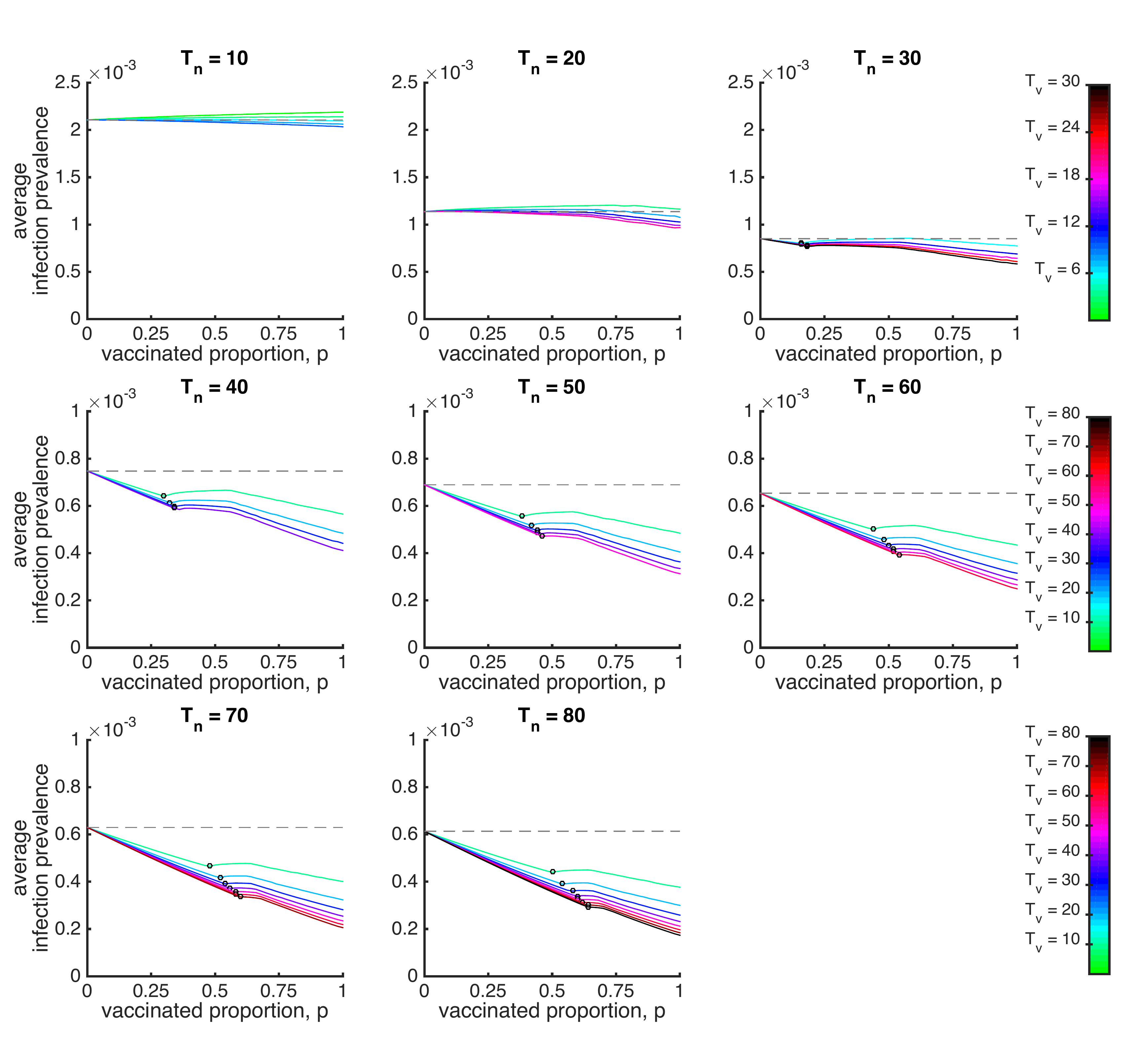}
	\caption{(Colour online) The mean infection prevalence for varying vaccinated proportion, duration of infection-acquired immunity (plots) and duration of vaccine-acquired immunity (colours). The gray dashed line indicates the mean infection prevalence in the absence of vaccination ($p = 0$). The black circle on each line denotes the Hopf bifurcation point that generates periodic cycles. ($1/\mu = 100$ years)} \label{fig:avg_i_prevalence_mu100}
\end{figure}

In this appendix, we repeat the numerical calculations for the system where the boosting of vaccine-acquired immunity bypasses primary infection (see Equation~\eqref{sirws}), but with a different average life expectancy. The average infection prevalence is calculated for an average life expectancy of 50 years (Figure~\ref{fig:avg_i_prevalence_mu50}) and 100 years (Figure~\ref{fig:avg_i_prevalence_mu100}). 

\newpage

\section*{References}

\bibliographystyle{elsarticle-harv} 
\bibliography{library}

\end{document}